**PassAI: explainable artificial intelligence algorithm for soccer pass analysis using multimodal information resources**

Ryota Takamido[1], Jun Ota[1], Hiroki Nakamoto[2]

[1] Research into Artifacts, Center for Engineering (RACE), School of Engineering, The University of Tokyo, 7-3-1 Hongo Bunkyo-ku, Tokyo, 113-8654, Japan

[2] Faculty of Physical Education, National Institute of Fitness and Sports in Kanoya, Kanoya, Kagoshima 891-2393, Japan

*Corresponding author:
Ryota Takamido
Email: ryota.takamido@gmail.com

Word count: 4750

**Abstract**

This study developed a new explainable artificial intelligence algorithm called PassAI, which classifies successful or failed passes in a soccer game and explains its rationale using both tracking and passer's seasonal stats information. This study aimed to address two primary challenges faced by artificial intelligence and machine learning algorithms in the sports domain: how to use different modality data for the analysis and how to explain the rationale of the outcome from multimodal perspectives. To address these challenges, PassAI has two processing streams for multimodal information: tracking image data and passer's stats and classifying pass success and failure. After completing the classification, it provides a rationale by either calculating the relative contribution between the different modality data or providing more detailed contribution factors within the modality. The results of the experiment with 6,349 passes of data obtained from professional soccer games revealed that PassAI showed higher classification performance than state-of-the-art algorithms by >5% and could visualize the rationale of the pass success/failure for both tracking and stats data. These results highlight the importance of using multimodality data in the sports domain to increase the performance of the artificial intelligence algorithm and explainability of the outcomes.

# 1. Introduction

The recent development of tracking technology has enhanced the understanding of complex athlete behavior in the real game context (Torres-Ronda et al., 2022). Because it provides a large datasets that include players' positions and velocities, accelerations, and other indexes (Giancola et al., 2018), analyzing these datasets has the potential to find new insights from the "data-driven" aspects (Fujii, 2021). In contrast, the creation and analysis of large datasets containing player performance indexes over games or entire seasons (i.e., "stats") have become increasingly common across many professional sports (Horvat & Job, 2020). Therefore, extracting meaningful information for practitioners from these vast and multimodal datasets is a key challenge in contemporary sports science, as traditional statistical analysis has difficulty in managing such complex datasets (Gandomi & Haider, 2015).

From a technical viewpoint, artificial intelligence or machine learning algorithms such as convolutional neural networks (CNN) (Yamashita et al., 2018) and recurrent neural networks (RNN) (Yu et al., 2019) have the advantage of processing large datasets consisting of noisy and highly complex data. However, several key issues still need to be addressed to effectively use artificial intelligence algorithms for sports analysis and coaching with big data.

The first challenge involves effectively using multimodal datasets for analysis. Multiple interacting factors influence sports performance. For example, in soccer pass behavior, both team tactics—such as the positions of other players recorded in the tracking data—and individual skill—such as the pass success rate captured in statistical data—can affect the outcome. If teammates create a wide "open space" (Fernandez & Bornn, 2018), pass success can be estimated independently of the passer's identity. However, in constrained spaces with limited passing options, success or failure depends on the passer's attributes. From a practical perspective, identifying the critical factors that influence performance outcomes is essential. However, most of the current artificial intelligence and machine learning algorithms in sports analysis were designed for unimodal datasets, making it difficult to analyze multimodal datasets despite their potential to enhance algorithm performance. This limitation may stem from the increased complexity of model and network architectures required to process multimodal datasets (Baltrušaitis et al., 2019).

The second challenge concerns the communication results to practitioners, specifically ensuring the explainability of artificial intelligence algorithm outcomes. Owing to the complexity and black-box nature of artificial intelligence computation, interpreting results remains difficult (Goodwin et al., 2022). Although several explainable artificial intelligence (XAI) techniques have been developed to clarify the rationale behind algorithmic outcomes (Minh et al., 2022), which have been applied by previous studies (Anzer et al., 2022), these efforts have primarily focused on unimodal data. Furthermore, the difficulty of generating explanations increases when multiple factors influence artificial intelligence outcomes (Rodis et al., 2024). Therefore, an additional underlying challenge is how to produce useful explanations for practitioners when using machine learning models that incorporate multimodal datasets.

## 2. Materials and Methods
### 2.1 Aim and contributions

Based on the above background, this study aims to address the two key issues described in the introduction. To that end, the success/failure pass classification using different modality datasets of tracking and stats data was set as the representative target task because (1) from the practical standpoint, the success or failure of a pass is a crucial factor that has a large impact on the outcome of the match (Adams et al., 2013), and how to organize the effective PassAIwork is one of the key aspects of the team dynamics (Garrido et al., 2020; Yamamoto & Yokoyama, 2011). (2) From a technical perspective, both spacing (positioning) information (Fernández & Bornn, 2021; Narizuka et al., 2021) included in tracking data and passer attribute (skill) information (Soniawan et al., 2021) can substantially affect pass success or failure.

This study developed a new XAI algorithm, PassAI, designed for soccer pass analysis. PassAI classifies the success or failure of specific passes made in real-game contexts by integrating multimodal data from tracking and seasonal stats data and provides explanations from multimodal perspectives. The contribution of this study lies in addressing aspects of machine-learning algorithm applications in sports science that have been insufficiently explored in previous studies, thereby enhancing their potential applicability in practical settings. Furthermore, from a narrower perspective, although many studies have been conducted to build a model to well predict the soccer pass success/failure probability and explain the rationale of it (Anzer & Bauer, 2022; Arbues-Sanguesa et al., 2020; Fernandez & Bornn, 2018; Gudmundsson & Wolle, 2012; Narizuka et al., 2021; Rahimian et al., 2023; Robberechts et al., 2023; Stöckl et al., 2021; Szczepański & McHale, 2016), no existing models can provide the explanation from multimodal perspectives. Therefore, the analysis of this study has the potential to provide new insights about the fundamental but important subjects of "Why is this specific path a success or failure?"

### 2.2 The proposed algorithm: PassAI
#### 2.2.1 Overview

Figure 1 presents an overview of the proposed PassAI algorithm. The task of PassAI is to classify the success/failure of a specific pass performed in the real game context based on the different modality data of (1) the tracking data, including the position, velocity, and acceleration of each player at the moment the pass occurred, and the ball departure and arrival position information, and (2) the seasonal stats of the passer, which represents the stats related to the pass action (e.g., total passes of the passer in the season). Specifically, to focus on the critical pass success/failure that contributes to scoring, the pass that arrives within the 30 m area of the goal is set as the target of the analysis.

The architecture of PassAI has two input streams, the ConvNeXt-Tiny module (Liu et al., 2022) and the multilayer perceptron (MLP) module (Cybenko, 1989) for processing the different modality inputs. The tracking data is represented by an image capturing the positions and velocities of each player at the moment the target pass is made. The stats information is represented as a feature vector, incorporating various passer-specific statistics, such as the total number of passes during the season. Classification was performed by integrating the outputs from both streams and processing them using a fully connected (FC) layer and a softmax function.

After determining the success or failure of the target pass, PassAI provides a rationale for the classification. PassAI employs a two-stage explanation process to generate explanations from a multimodal perspective. In the first stage, PassAI visualizes the relative contributions of the tracking and stats data to classify the target pass based on a gradient calculation of the input features. This explanation provides rough feedback on "which information was given more weight by the artificial intelligence algorithm." In the second stage, PassAI identifies the most influential factors within each modality. For the tracking image, key area information in the input image by highlighting the contributing parts using Gradient-weighted Class Activation Mapping (Grad-CAM) (Selvaraju et al., 2017). For the stats data, the contributions of each feature are visualized by gradient calculations. Through these two explanations, users estimate the potential reasons for the success/failure of a specific pass made in a real game from both positionings and the attributes of the passer. Further technical details regarding PassAI and its breakthroughs are discussed in the following sections.

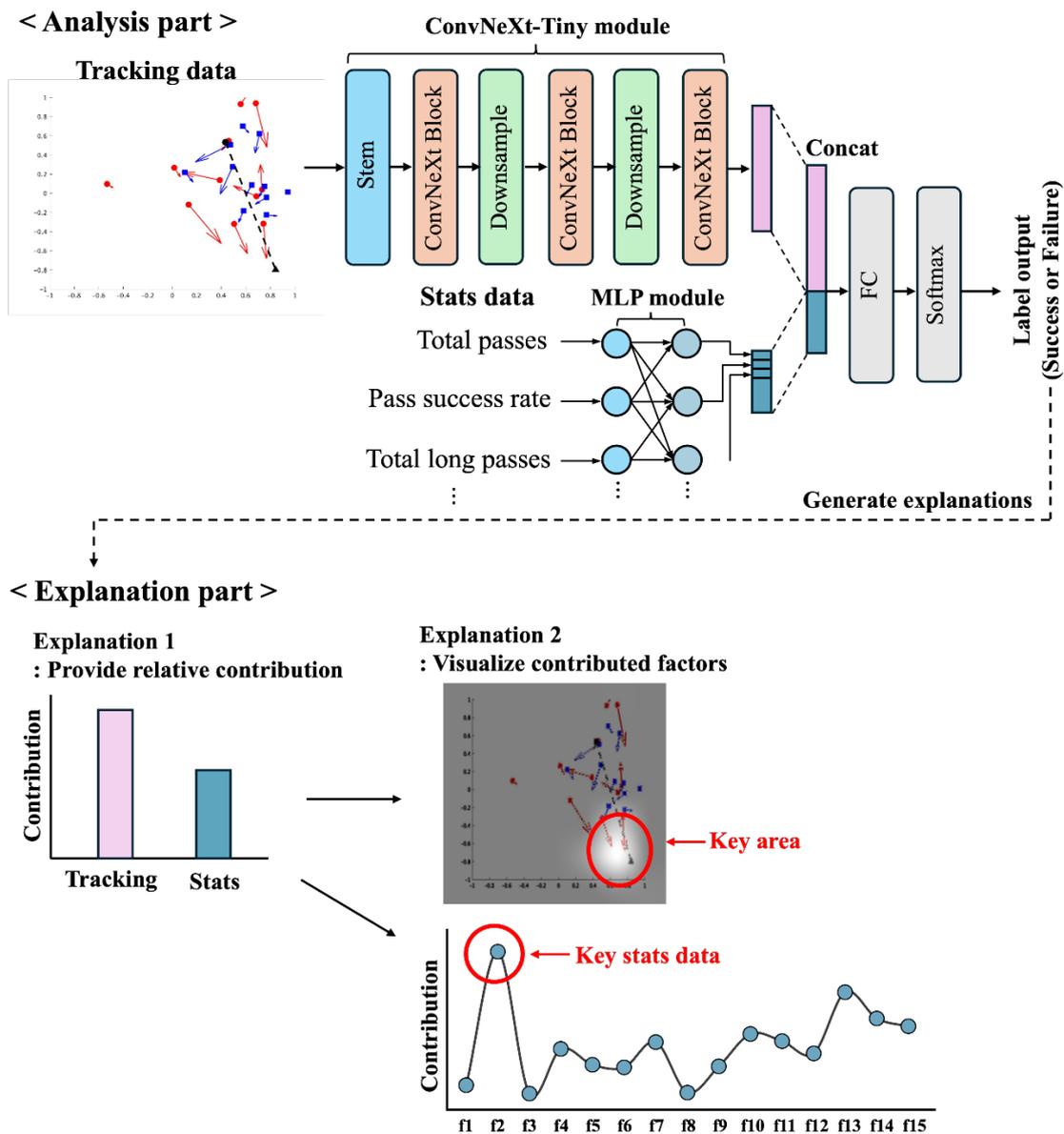

Figure 1. The overview of the proposed machine learning algorithm: PassAI.

### 2.2.2 Input data

PassAI uses two different modality inputs for the network input data. One is tracking image data that captures the behavioral information of the players and balls at the moment when the target pass occurs. The other is the feature vector, which represents the stats of the passer related to the passing performance, such as the total pass rate and success rate of the passer during the season. To the best of our knowledge, this study is the first study to process both tracking and stats data using an XAI algorithm and visualize the outcome rationale.

Specifically, Figure 2 shows the details of the image input. As shown in the figure, the image includes various types of information that can potentially affect pass failure/success, such as the position and velocity information of each player and the ball position when the pass and next play occur (e.g., receiving or intercepting). Although most previous studies used graph representations that map each player's information to nodes and establish edges between interacting players (Cartas, Ballester, & Haro, 2022; Wang et al., 2024; Xenopoulos & Silva, 2021), this study specifically adopted an image-based representation. This approach offers the advantage of explicitly conveying open-space information (Battaglia et al., 2018) by representing these areas with RGB (red-green-blue) values of [255 255 255] (i.e., white color). The size of the field is scaled, and the values of -1 and 1 for the *x* and *y* axes indicate both ends of the field. Offensive and defensive teams are distinguished by color, and the departure and arrival points of the ball are connected by a dotted line.

Furthermore, PassAI uses the input of the passer's stats information, which is represented by a 15-dimensional feature vector consisting of the indexes listed in Table 1. These features provide significant information about the type of player who made the pass. For example, the number of long passes tends to be higher for defenders and lower for forwards, and a large number of total passes suggests that the passer plays an important role as a hub in the passing network (Yamamoto & Yokoyama, 2011). To account for the bias in success rates when the total number is extremely small, the product of the total number and success rate is also inputted as features. The value of each feature was standardized to a normal distribution for all passers.

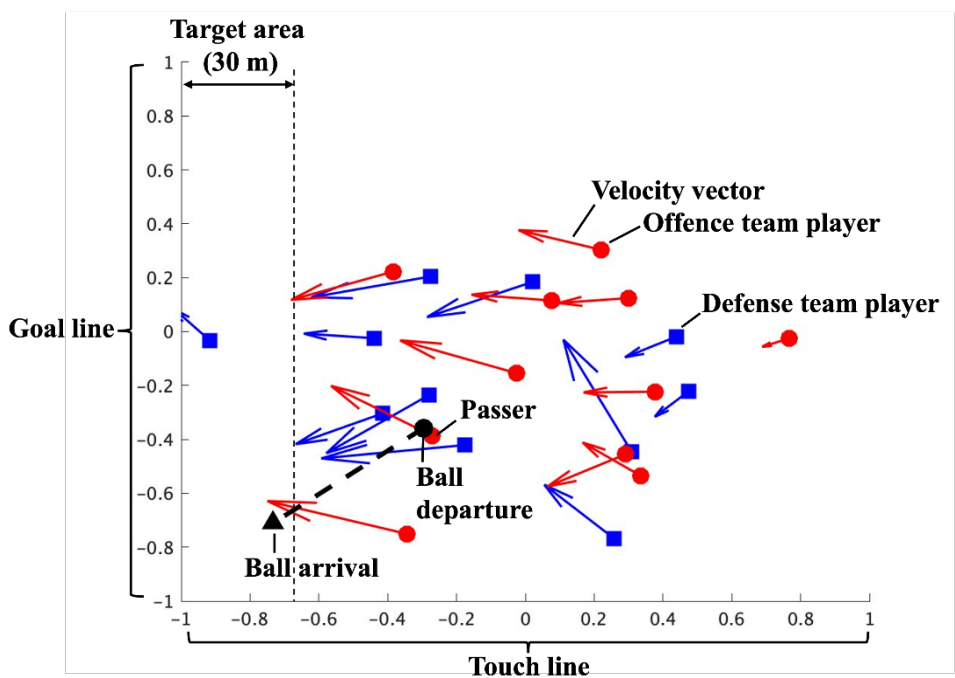

Figure 2. The tracking image data used as the input to PassAI.

Table 1. The list of the passer's stats information used in the input feature vector.

| Number | Feature | Brief explanation |
|---|---|---|
| 1 | Total passes | The total number of all passes made by the player through the target season. |
| 2 | Pass success rate | The success rate of passes made by the player. |
| 3 | Total passes × pass success rate | Multiplication of variables 1 and 2. This can enable us to consider the interaction effect between these two variables. |
| 4 | Total passes for opposition area | The total number of passes made by the player that arrived at the opposition area. |
| 5 | Pass success rate for opposition area | The success rate of the above-typed pass. |
| 6 | Total passes for opposition area × Pass success rate for opposition area | Multiplication of variables 5 and 6. |
| 7 | Total long passes | The total number of long passes that travel more than 30 m |
| 8 | Long pass success rate | The success rate of the above-typed pass. |
| 9 | Total long passes × Long pass success rate | Multiplication of variables 7 and 8. |
| 10 | Total through passes | The total number of through passes aimed to allow teammates to run onto the ball and create a scoring chance. |
| 11 | Through pass success rate | The success rate of the above-typed pass. |
| 12 | Total through passes × Through pass success rate | Multiplication of variables 9 and 10. |
| 13 | Total cross passes | The total number of cross passes sent to the penalty area from the outside. |
| 14 | Cross pass success rate | The success rate of the above-typed pass. |
| 15 | Total cross passes × Cross pass success rate | Multiplication of variables 11 and 12. |

**2.2.3 Network architecture**

Because the image information had a larger data size than the feature vector (224×224×3 vs. 15×1), it was processed separately in two modules with different levels of complexity (Figure 1). In this study, ConvNeXt was selected to process the image data (Liu et al., 2022), whereas previous studies used GoogleNet (Szegedy et al., 2015) for the corresponding part (Wagenaar et al., 2017; Raabe et al., 2023). ConvNeXt is a state-of-the-art CNN-based algorithm. It is inspired by the Vision Transformer (ViT) (Dosovitskiy, 2020), which has the advantage of capturing the interaction between pixels in distant positions in the image, and improves traditional CNN by incorporating design elements from ViT while retaining the advantage of CNN-based architecture to convolute the local information deeply. Therefore, ConvNeXt is expected to effectively process both cases where the players are sparsely arranged and make long-distance interactions (e.g., long pass) and where the players are densely arranged and make short-distance interactions (e.g., short pass near the goal) and to show a higher classification performance than the other image processing algorithms.

ConvNeXt-tiny (Liu et al., 2022) was used to balance the accuracy and processing speed of the algorithm. In the model (Figure 1), the 224 × 224 input image was first processed by the stem cell with a patchify layer implemented using a 4×4 stride 4 convolutional layer. The output from the stem cell was then processed by the ConvNeXt block, consisting of a 7×7 depthwise convolution layer, layer normalization, and a 1×1 convolution layer. The larger kernel size of the depthwise convolution layer contributes to expanding the receptive field of the convolution layer and captures long-distance interactions in the image (Ding, Zhang, Han, & Ding, 2022). Downsampling is performed using 2×2 conv layers with stride 2 after the ConvNeXt block. Finally, outputs of 768×1 features were obtained from this stream. For more technical details, please refer to the original paper (Liu et al., 2022).

Regarding the stats information, the feature vector was processed using a multilayer perceptron (MLP) (Cybenko, 1989) with one hidden layer having 64 neural units. The MLP is a type of feedforward neural network that contains at least three layers of neurons (input, hidden, and output layers) (Neagoe, Ciotec, & Cucu, 2018). The neural units in the hidden and output layers have a nonlinear activation function for modeling complex interactions among the input features. By processing the feature vector using the MLP module, PassAI can handle the complex interactions among the indexes in Table 1, such as players who have relatively larger long pass numbers but smaller pass numbers (e.g., a defender). Consequently, the outputs of 64×1 features were obtained from this stream. The output of the MLP module is combined with that of the ConvNext-tiny module; the combined data are input to the fully connected (FC) layer and softmax function, and the label for the target pass (success/failure) is generated.

**2.2.4 Rationale for classification**

After outputting the label of the target pass, the rationale is explained to users. To generate explanations from a multimodal perspective, a two-stage explanation was performed by PassAI to provide information on (1) which modality data and (2) features received more weight to analyze the target pass.

In the first stage, the relative contributions of the tracking and stats data were determined by gradient (sensitivity) calculation with respect to the output (Figure 1). For example, if the classification is heavily based on the contextual image, the gradient for each pixel will be large. However, in cases where the gradient of the feature vector is large, the success or failure of a pass depends more on the attributes of the passer.

Specifically, given $y^c$ is the output from the FC layer and the $x_{i,j,k}$ represent the pixel value for height $i$, width $j$, and channel $k$ in the image, the relative contribution of the tracking image data $C_T$ can be described by the following equation:

$$C_T = \sum_i \sum_j \sum_k \frac{\partial y^c}{\partial x_{i,j,k}}. \qquad (1)$$

Similarly, given the value of the *i*-th feature vector $v_i$, the relative contribution of passer's stats information can be described by the following equation:

$$C_S = \sum_i \frac{\partial y^c}{\partial v_i}. \qquad (2)$$

Although directly comparing $C_T$ and $C_S$ is challenging, comparing $C_T$ ($C_S$) for a specific pass with $C_T$ ($C_S$) for another pass is possible. Therefore, this analysis provides insights such as "the passer information is relatively dominant in this pass classification than the former pass." The value of $C_T$ and $C_S$ is standardized within 0–1 for all passes in the analysis.

Subsequently, a more detailed explanation for each modality is generated in the second stage. For the tracking image data, PassAI generates an image to highlight the key areas in the field using Grad-CAM (Selvaraju ET AL., 2017) (Figure 1). Grad-CAM is a representative explanation technique for image recognition that visualizes the contribution of the input in any convolutional layer using gradients for the target class. In this study, we applied Grad-CAM to the last convolutional layer of ConvNext to highlight the areas in the image that significantly contributed to the prediction of pass success/failure (Figure 1).

Specifically, in the Grad-CAM process (Selvaraju et al., 2017), the localization map $L^c_{Grad-CAM} \in \mathbb{R}^{u \times v}$ of width $u$ and height $v$ for the target class $c$ is calculated by the following equations:

$$\alpha^c_k = \frac{1}{Z} \sum_i \sum_j \frac{\partial y^c}{\partial A^k_{ij}}, \qquad (3)$$

$$L^c_{Grad-CAM} = RELU\left(\sum_k \alpha^c_k A^k\right), \qquad (4)$$

where $Z$ is the number of the elements in the feature map $A$ and weights $\alpha^c_k$ represents the partial linearization of the network downstream from $A$. The activation function (*RELU*) in Equation (4) converts negative values to zero, and a heat map for highlighting the image can be obtained.

For the passer's stats data, PassAI identified the most influential factors within the stats information (Table 1) by calculating the gradients of each feature variable obtained during the computation of equation (2). This approach allows users to determine which passer attributes are mostly used in the classification of the target pass (Figure 1).

### 3. Experiment with real game data
### 3.1 Datasets

This study uses data from 95 games in the 2023 and 2024 seasons of the Japan professional Football League (J1). This dataset was provided by Data Stadium, Inc. The dataset included tracking, event, and player data. The tracking data included the positions of the players and the ball at each time frame of 25 Hz. The event data include the label of actions, such as the pass, the time frame when the action occurred, and its result (success/failure). Player data represents the names and teams of each player. The company was authorized to obtain these data and sell them to third parties, and it was guaranteed that the use of the data would not infringe on the rights of players or teams. Additionally, from the official website of the J1 League (https://www.jleague.jp/), the stats information included in Table 1 was collected for each passer. Although the dataset is not publicly shared owing to rights issues, the authors are willing to provide detailed explanations upon request.

To generate the training and test datasets, the passes that arrived inside the 30 m area from the goal were extracted. As a result, 3,663 successful and 2,686 failed passes were selected from the original data. Subsequently, the position and velocity of each player at the time when the pass occurred and the ball position at the time when the pass and next action occurred were collected. The tracking image data were generated for each pass using the collected data.

### 3.2 Model training and verification test

After generating the dataset, PassAI was trained, and the performance of the pass success/failure classification was verified. First, the generated datasets were divided into training, validation, and test sets at a ratio of 8:1:1. Since this study adopted the 10-fold cross-validation method (Bradshaw, Huemann, Hu, & Rahmim, 2023), the partition between the datasets was changed for every 10 validations. When training was performed, data augmentation was randomly applied to reverse the left and right and the top and bottom of the image by referencing the method of a previous study (Wang et al., 2024). The implementation was performed on an NVIDIA v2-8 TPU with Tensor Core on a Python platform. To reduce the training time, a pretrained ConvNext-tiny model was installed, and fine-tuning was performed. The batch size was set to 128, the maximum number of epochs was set to 10, and the learning rate was set to 1e-4. The model was trained to minimize the summation of cross-entropy losses using the Adam optimization solver, and the model that showed the highest accuracy for the validation datasets during the training process was saved for testing.

Furthermore, to make state-of-the-art comparisons with existing algorithms, some existing algorithms were also implemented. Specifically, this study used GoogleNet (Szegedy et al., 2015),

ResNet50 (He et al., 2016), EfficientNet-v2 (Tan & Le, 2021), and ViT (Dosovitskiy, 2020) as the benchmark image-based methods to compare the classification performance with ConvNeXt. The training was conducted using only tracking image information. While only GoogleNet has been used as the benchmark algorithm for image-based methods in previous research (Raabe et al., 2023), this study conducted a more comprehensive verification. Additionally, to verify the effect of multimodal information use on the classification performance, we implemented the Pure-ConvNext model, which was trained using only the tracking image information. The training and evaluation procedures for these models were identical to those for the proposed method.

The graph convolutional neural network (GCN) and Granger causality-inspired graph neural network (CI-GNN) (Zheng et al., 2024) were implemented as graph-based methods to verify the effect of the tracking data. The basic GCN architecture design was based on a previously described method (Raabe et al., 2023). Note that, while the method of the previous study first processed the time series data using a Gated Recurrent Unit (GRU) (Chung et al., 2014) and then convolved it using a GCN, this study partially referenced the GCN part owing to the use of the non-time series input in our task. The CI-GNN is a state-of-the art explainable graph neural network inspired by the concepts of Granger causality (Granger, 1969) and variational graph autoencoders (Kipf & Welling, 2016). It extracts and highlights the important local structures that have causal relationships with the classification results. For a fair comparison with the graph-based algorithms in our task, we added new information on the distance between each player's position and the ball's departure/arrival positions as the node feature.

## 4. Results
### 4.1 Prediction performance

Table 2 lists the pass success/failure classification performance of each algorithm. As shown in the table, PassAI exhibited the highest performance for all evaluation indexes. It can classify the pass successes/failures with an accuracy of 77.6%. Figure 3 shows the confusion matrix for the PassAI classification. From the figure, PassAI has a better prediction accuracy for the successful passes (83.0%) than for failed passes (70.8%). This may be due to an imbalance in the class (3,663 successes and 2,686 failures).

Regarding the performance of other existing algorithms, ViT and ConvNeXt performed better than the other image-based methods. This suggests the advantage of the transformer-based architecture, which enables the model to capture long-distance interactions in an image. From a comparison with PassAI and Pure-ConvNeXt models, the use of multimodal input data can increase the performance indexes by 2–4%. Furthermore, the graph-based method also showed values of approximately 70% for each evaluation index. The CI-GCN had a relatively higher F1 score and recall because of its high classification accuracy for the minor class (failed passes).

Table 2. The pass classification performance of implemented algorithms.

|  |  | Accuracy | Precision | Recall | F1-score |
|---|---|---|---|---|---|
| Image-base | GoogleNet | 0.550 | 0.550 | 0.308 | 0.394 |
|  | ResNet-50 | 0.667 | 0.666 | 0.666 | 0.662 |
|  | EfficientNet-V2 | 0.603 | 0.602 | 0.600 | 0.600 |
|  | ViT | 0.694 | 0.692 | 0.694 | 0.694 |
|  | Pure-ConvNeXt | 0.753 | 0.756 | 0.735 | 0.740 |
|  | **PassNet (Proposed)** | **0.776** | **0.780** | **0.780** | **0.778** |
| Graph-base | GCN | 0.697 | 0.693 | 0.683 | 0.684 |
|  | CI-GNN | 0.663 | 0.688 | 0.719 | 0.758 |

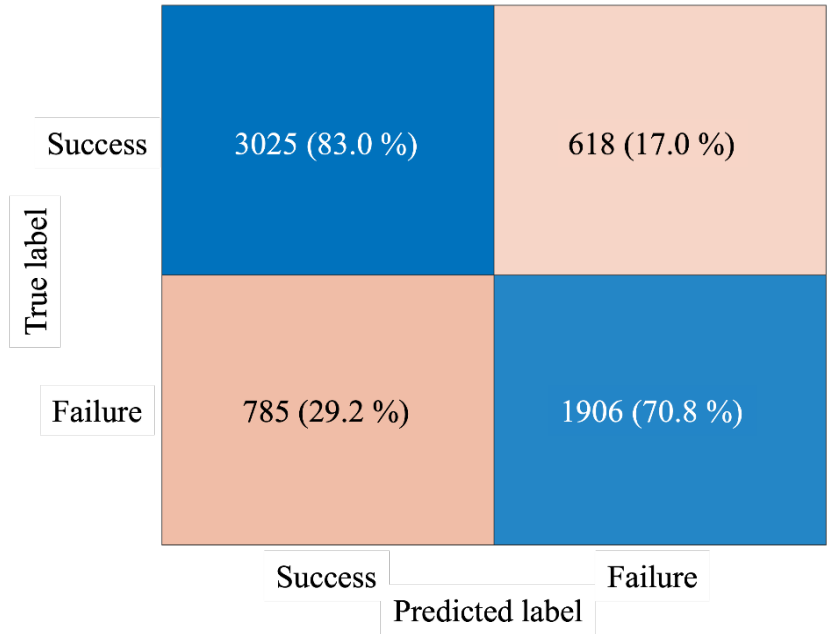

Figure 3. Confusion matrix generated from the PassAI classification results. The values in parentheses represent line-normalized probabilities.

## 4.2 Explanation results

Figure 4 shows the calculated relative contributions for all the data provided by the first-stage explanations. As shown in the figure, the contribution of the tracking and stats data changed among passes. This finding suggests that PassAI alters the data primarily used for classifying each pass.

Figure 5 shows the tracking images for the cases where the image input contributed more (Figures 5 (a) and (b)) and the passer's stats contributed more (Figures 5 (c) and (d)). All four cases in

Figure 5 represent correct prediction cases for a successful pass. In cases where the image input contributes more (Figures 5 (a) and (b)), the direction of the speed vector of the teammate matches the direction of the ball, and the influence of this contextual information is given more weight than the passer attribution. Meanwhile, PassAI seems to place more weight on passer information when it is difficult to make a judgment based only on the image (Figure 5 (c) and (d)). For example, in the case of Figure 5 (c), several players near the ball arrival position are highly competitive, and success/failure can change owing to slight fluctuations such as in the speed of the ball. In the case of Figure 5 (d), the potential receiver near the ball arrival position has a velocity vector that is different from the direction of the ball; success/failure also depends on the nature of the pass (e.g., a pass that is too fast does not provide sufficient time to change the direction of the receiver).

Regarding the explanations performed by the second stage, Figure 6 shows an example of the Grad-CAM images, highlighting the key areas for pass success/failure classification. From the figure, we can find different factors that influence pass success/failure classification, such as the open space near the pass arrival position (Figure 6 (a)), motion of the key player (Figure 6 (d)), density near the pass arrival location (Figure 6 (c)), and more global player arrangement (Figure 6 (b)). Thus, by visualizing the important parts of the image, in addition to the success/failure classification, the user can obtain detailed explanations of its rationale.

Figure 7 presents the second-stage explanation for the passer's stats information. The factors contributing to classification varied for each pass, allowing users to identify which parameters played a more significant role in a given case (Figure 7 (a)–(b)). Figure 7 (c) represents the average contribution of each stats feature across all passes. Notably, the seasonal pass success rate of the passer was a major contributing factor (Figure 7 (c), f2). Pass performances (Figure 7 (c), f11 and f12) exhibited high contributions, likely because of the tendency of midfielders—who often execute through passes—to possess strong passing skills.

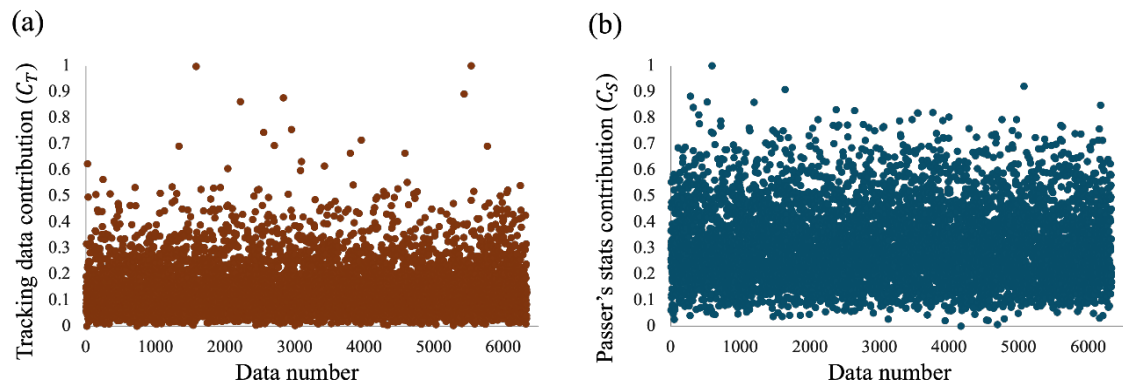

Figure 4. The calculated relative contributions of (a) tracking image data ($C_T$) and (b) passer's stats information ($C_S$) for all data.

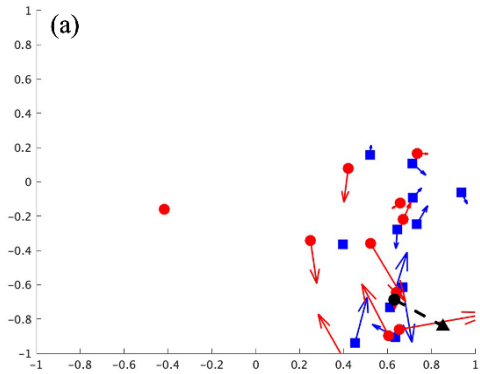
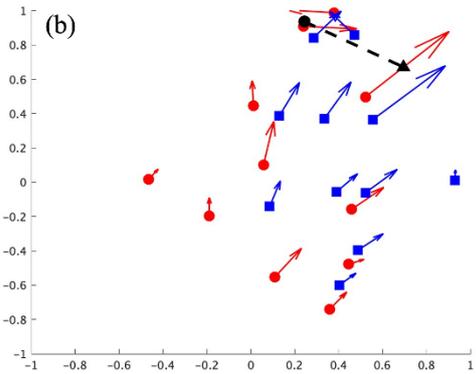
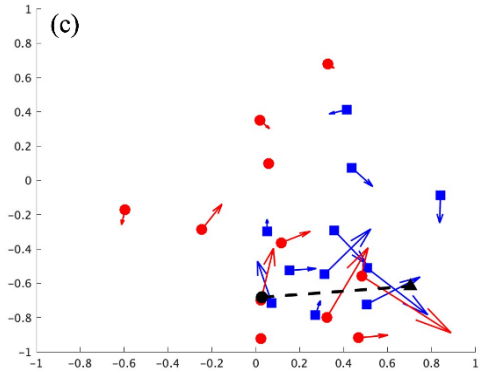
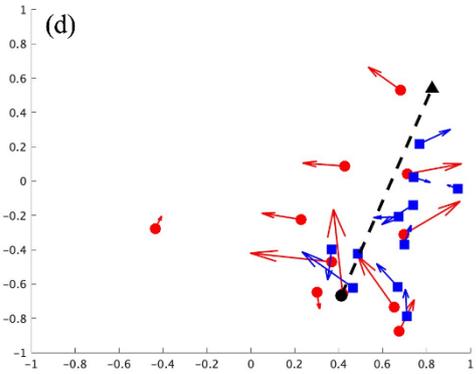

Figure 5. Example explanations obtained from PassAI. (a-b) Cases were tracking image information contribution (sensitivity) is relatively high ($C_T = 1.0, 0.89$, respectively), (c-d) Cases were passer's stats information contribution is relatively high ($C_S = 0.84, 0.82$, respectively). All four cases correspond to correctly predicted successful pass.

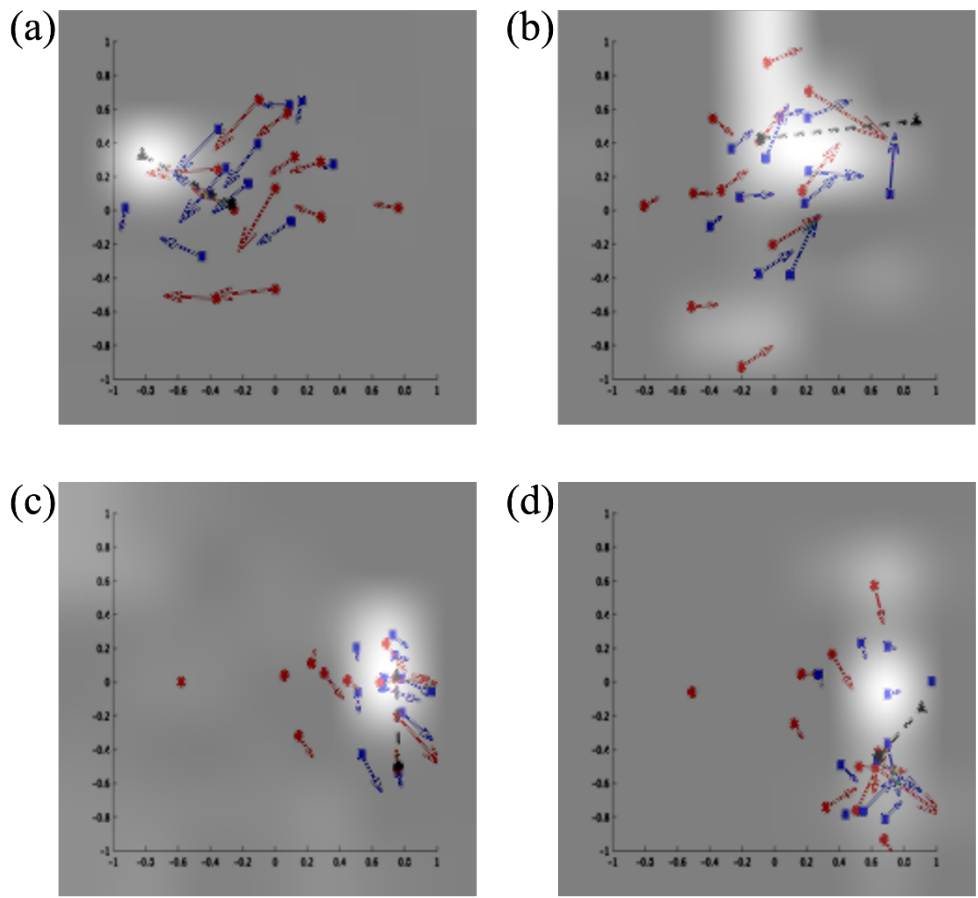

Figure 6. The example of the Grad-CAM images obtained from PassAI. (a-b) Explanations for correctly predicted successful passes. (c-d) Explanations for failed passes.

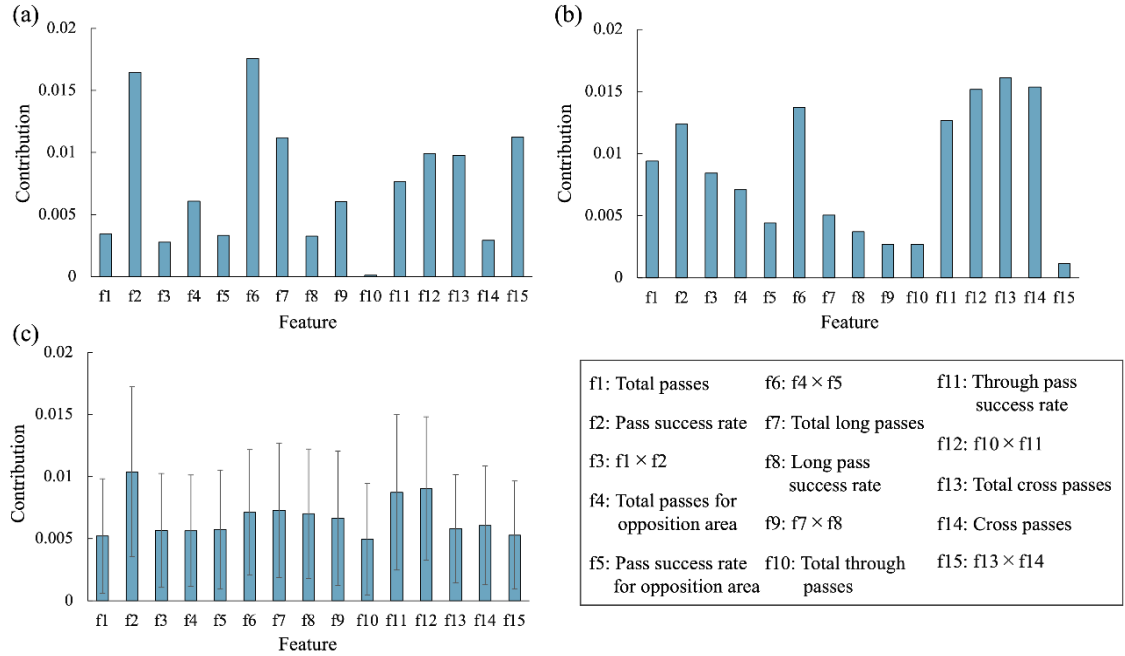

Figure 7. Analysis of contributions of each passer's stats information. (a–b) Example of the specific passes, (c) average values among all data.

5. Discussion

One notable finding is that the integration of both tracking and stats data significantly enhances classification performance (Table 2). Although previous studies primarily focused on optimizing the use of unimodal data, particularly mostly tracking data (Fernández & Bornn, 2021; Wagenaar et al., 2017), the incorporation of multimodal data provides substantial benefits to artificial intelligence algorithms. Because one of the unique characteristics of human behavior in the sports context is the individual difference among players (Müller et al., 2015; Takamido et al., 2019), considering both whole-field spatial dynamics (tracking image) and individual player attributes (passer's stats) may be broadly applicable across various sport-related tasks.

Additionally, the results of the comparison of classification performance showed that image-based methods outperformed graph-based methods in the task of this study. This suggests the importance of a suitable data representation based on the characteristics and requirements of the target sports actions. Although the graph-based representation has been widely used for the analysis of the team sports, image-based representation may be more suitable for the tasks that the open space information is contributed. For further improvements of the image-based method, it may be beneficial to add some "additional" information into the image, such as the pressure distributions from the defender (Andrienko et al., 2017) or the control space of each player (Raabe et al., 2024). Additionally, because the graph-based method also showed relatively high-performance indices (approximately

70%), the improvements in the current graph-based methods, such as extending the edges among players (Wang et al., 2022) or using a larger kernel size to capture the long-range interactions (Liu et al., 2023), are the other prominent work.

Furthermore, by incorporating the XAI technique, this study demonstrated the potential to visualize the rationale behind the success or failure of individual actions, as shown in Figures 4–7. Providing effective feedback to learners is a key issue in sports coaching (Moon, 2022). While previous studies have mainly used explanation techniques to identify important feature variables within the same modality data (Anzer & Bauer, 2022; Forcher et al., 2024; Wang et al., 2022), this study provides more comprehensive explanation methods. However, the explanatory methods adopted in this study require further refinement to better align with the specific characteristics of sports. For example, based on our observations, some generated visualizations were difficult to interpret, as shown in Figure 8. Addressing this issue may require training the network to generate better explanations based on the evaluations of each image by skilled coaches or integrating the image captioning technique (Hossain et al., 2019) to provide more concrete information. Additionally, the development of quantitative and concrete evaluation methods for generated explanations is essential (Lopes et al., 2022). These methods should be grounded in domain-specific knowledge to ensure their relevance and effectiveness in sports analysis.

From a practical standpoint, PassAI can be utilized more effectively by integrating it with a suggestion system, as explored in previous studies (Fernández & Bornn, 2021; Liu et al., 2024; Wang, Qin, & Liu, 2024). For example, if we change the ball arrival locations in the same image and analyze those images using PassAI, we can identify the areas with a higher probability of success. Additionally, if we change the positions of the players in the image and analyze those images, it provides a better formation to increase the probability of pass success (or intercept).

Finally, this study had several limitations. First, as mentioned earlier, the explanation methods used in this study require further refinement to better align with the sports domain. Future studies should build on these findings to develop more specialized explanation techniques tailored to sports applications. Second, regarding the network architecture, this study adopted the existing ConvNeXt and MLP models. Although these showed high performances, further improvements are necessary to enhance the effectiveness of PassAI. Finally, as this study used only professional league data, future studies should explore the applicability of the proposed methods across different skill levels. This is essential to promote the broader adoption of these technologies across various age groups and competition levels.

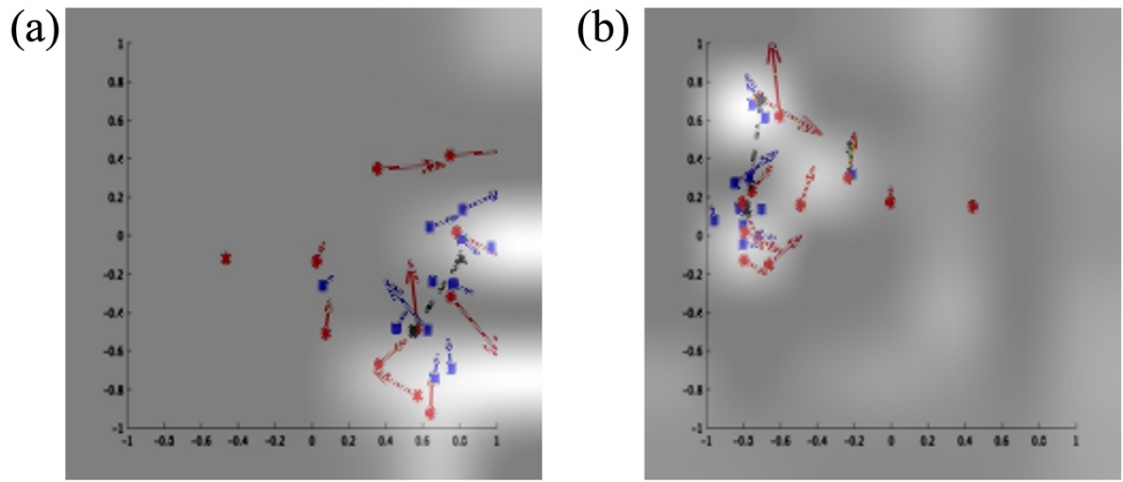

Figure 8. The example of the explanation images that are difficult to directly interpret.

## 6. Conclusion

In conclusion, the high classification performance of PassAI highlights the importance of incorporating multimodal data when analyzing sports performance. Although more sophisticated explanation methods should be developed in the future while considering the unique characteristics of sports coaching and education, the inclusion of a two-stage explanation module for interpreting the role of each modality provides valuable insights into the outcomes generated by PassAI. Based on the results of this study, future research is expected to advance knowledge on the effective use and development of XAI algorithms in each sport domain.